\documentclass[copyright,creativecommons]{eptcs}

\usepackage{amsmath}
\usepackage{graphicx}
\usepackage{multirow}

\usepackage{amssymb}
\usepackage{hyperref}
\usepackage{booktabs}
\usepackage[english]{babel}

\usepackage{graphicx}
\usepackage{mathptmx}
\usepackage{tikz-qtree}
\usepackage{filecontents}
\usepackage{csvsimple}
\usepackage{amsfonts,amsmath}
\usepackage{subfig} 
\usepackage{url}
\usepackage{verbatim}
\usepackage{color}
\usepackage{amsmath}
\usepackage{proof}
\usepackage{listings}
\usepackage{xcolor}
\usepackage{fancyvrb}
\renewcommand{\phi}{\varphi}

\input prolog.tex

\definecolor{lgray}{gray}{0.95}
\definecolor{lblue}{rgb}{0.90,0.90,1.00}
\definecolor{lyellow}{rgb}{1.00,1.00,0.70}

\newtheorem{prop}{Proposition}

\newtheorem{ex}{Example}

\DefineVerbatimEnvironment{codex}
{Verbatim}
{fontsize=\small, frame=single, formatcom=\color{blue}}

\newcommand{\BI}[0]{\begin{itemize}}
\newcommand{\EI}[0]{\end{itemize}}

\newcommand{\BE}[0]{\begin{enumerate}}
\newcommand{\EE}[0]{\end{enumerate}}

\newcommand{\BX}[0]{\begin{ex}}
\newcommand{\EX}[0]{\end{ex}}
\newcommand{\BV}[0]{\begin{verbatim}}
\newcommand{\EV}[0]{\end{verbatim}}

\newcommand{\BP}[0]{\begin{prop}}
\newcommand{\EP}[0]{\end{prop}}

\newcommand{\BEQ}{\begin{equation}}
\newcommand{\EEQ}{\end{equation}}

\newcommand{\BC}[0]{\begin{center}}
\newcommand{\EC}[0]{\end{center}}

\newcommand{\BF}[0]{\begin{filecontents*}{data.csv}}

\newcommand{\BQ}[0]{\color{blue}\begin{quote}}
\newcommand{\EQ}[0]{\end{quote}\color{black}}

\def \bscale1 {0.25}
\def \bscale {0.25}

\usepackage{lineno} 

\def\titlerunning{On the (Intuitionistic) Logic of Next-Token Prediction}
\def\authorrunning{P. Tarau}

\begin{document}


\title{On the (Intuitionistic) Logic of Next-Token Prediction}

\author{Paul Tarau
  \institute{University of North Texas}
  \email{Paul.Tarau@unt.edu}
  }


\maketitle

\begin{abstract}
We model in intuitionistic implicational logic the key enabler of today's GenerativeAI: the next-token prediction in autoregressive causal neural networks.

In our framework, next-token prediction corresponds to \emph{modus ponens}, and sequence processing becomes constructive proof extension under the Curry--Howard correspondence.
Our Prolog-based specialized theorem provers validate  fundamental properties of the neural models, among which relations between commutative vs. non-commutative sequencing  and single-token vs. multi-token prediction choices.

We derive a neural architecture equivalent to multiplicative RNNs
that arises naturally from a proof-theoretic interpretation of next-token
prediction as nested intuitionistic implication
and position the model relative to transformers, state-space models and recursive LLMs.

\end{abstract}

\section{Introduction}

Foundational Generative AI models involve  mechanisms like next token(s) prediction
that naturally benefit
from unsupervised training on very large data sets
at a scale closely matching the total multi-modal content of the internet.

The resulting distribution, encapsulated in the parametric memory of
the model, {\em constructively} completes the initial sequence
derived from the prompt and subsequently generated tokens with
a set of probability-weighted next token candidates.

The underlying logic of the derivation of the next token
hints to an intuitionistic implicational logic  proof of  $q$ seen as an implication from assumptions in context $E$,
$E \rightarrow q$.
This is can be stated as
{\em finding a way to transform a proof of an assumption context $E$ into a proof of
    a consequence $q$}.
In the neural world this materializes as
a transformation of the overall statistical evidence for
$E$ into a predicted probability for $q$.

In the case of the transformer models \cite{transfo},
order information is learned via dedicated positional encodings
that complement the parallel attention heads that learn
deep dependencies relating to the ``past'' token sets
while having the set of ``future tokens'' masked.
 This is a highly effective approach, but it obscures the logical structure of sequence prediction and provides little explanatory connection to symbolic reasoning.

We will explore here an alternative view: next-token prediction as implication completion. We ask whether the task of predicting the next word in a sequence can be understood as extending a proof in intuitionistic propositional logic—and whether this interpretation leads naturally to a viable neural architecture.

Our answer is affirmative. Starting from a symbolic encoding of sequences as left-nested intuitionistic implications, we derive a recurrent neural model in which each token acts as an operator transforming a proof state. The resulting architecture, which we call the {\em Arrow Language Model}, replaces similarity-based attention with non-commutative operator composition as the carrier of both semantic content and order.

Our logic-based insights will also be extended to cover multi-token prediction
and it naturally relates to type inference and type inhabitation via the Curry-Howard isomorphism.

The rest of the paper is organized as follows.
Section \ref{tools} introduces our theorem provers for propositional implicational intuitionistic logic.
Section \ref{intu} justifies our choice of the logic formalism modeling content and order
in the next token prediction.
Section \ref{theo} illustrates instances of valid formulas relating left-nested implicational sequences,
relevant  for the next word(s) prediction task.
Section \ref{info} introduces an information retrieval mechanism by using implicational logic subformulas
built  from a corpus split into sentences.
Section \ref{neural} describes the a neural realization of the derived {\em Arrow Architecture}.
Section \ref{exper} describes our experiments involving validation of our logic-based algorithms and their neural equivalents.
Section \ref{disc} discusses limitations and future work directions.
Section \ref{rel} overviews related work.
Section \ref{conc} concludes the paper.

Our full open-source code\footnote{ \url{https://github.com/ptarau/nextword}}
and the literate Prolog code extracted from this
paper \footnote{\url{https://github.com/ptarau/nextword/blob/main/nextword.pro}}
are available online.

\section{Our tools :  implicational intuitionistic logic theorem provers}\label{tools}

We will start by motivating  our choice of a purely symbolic logic representation that does  not involve reliance on numerical implementations of ordering.

Implicational propositional logic allows reasoning, when seen as
a Hilbert system with modus-ponens as its only inference rule:
\[
\begin{gathered}
    p \to q \\
    p \\
    \hline
    \therefore q
\end{gathered}
\]
and two axioms corresponding via the Curry-Howard isomorphism
to combinators K and S:
\begin{codex}
K combinator: p->q->p

S combinator: (p->q->r)->(p->q)->p->r
\end{codex}

For the same formalism, when seen via the sequent calculus, simple and efficient theorem provers
become available.
We will introduce them next as they will be used to support our logical inference mechanisms.
We implement Roy Dyckhoff's {\bf LJT} calculus, restricted to the implicational fragment of propositional intuitionistic logic, proven as sound and complete in \cite{dy1}, while we also construct the proof represented as a $\lambda$-term.
\medskip

{\large
    \noindent
    \begin{math}
        LJT_1:~~~~\frac{~}{A,\Gamma ~\vdash~ A}\\\\
        LJT_2:~~~~\frac{A,\Gamma ~\vdash~ B}{\Gamma ~\vdash~ A\rightarrow B}\\\\
        LJT_3:~~~~\frac{B,A,\Gamma ~\vdash~ G}{A \rightarrow B,A,\Gamma ~\vdash~ G}\\\\ 
        LJT_4:~~~~\frac{D \rightarrow B,\Gamma ~\vdash~ C \rightarrow D ~~~~ B,\Gamma ~\vdash~ G}
        { \left( C \rightarrow D \right) \rightarrow B,\Gamma ~\vdash~ G }\\
    \end{math}
}
~\\
Termination is ensured as one can identify a multiset
ordering-based size definition that decreases after each step \cite{dy1}.

The predicate {\tt lprove/2}'s proof term is built by introducing
$\lambda$-terms {\tt l(X,E)} with $\lambda$-variable {\tt X} and expression {\tt E},
and  representing application of a source {\tt A} to  target {\tt B} as {\tt a(A,B)}.
The predicate {\tt iprove/1} ignores the last argument of {\tt lprove/2} for cases when that is not needed.
\begin{code}

lprove(T,X):-lprove(X,T,[]),!.

iprove(T):-lprove(T,_).

lprove(X,A,Vs):-memberchk(X:A,Vs),!. 
lprove(l(X,E),(A->B),Vs):-!,lprove(E,B,[X:A|Vs]).  
lprove(E,G,Vs1):-
    select(S:(A->B),Vs1,Vs2),       
    lprove_imp(T,A,B,Vs2),          
    !,
    lprove(E,G,[a(S,T):B|Vs2]).     

lprove_imp(l(X,E),(C->D),B,Vs):-!,lprove(E,(C->D),[X:(D->B)|Vs]).
lprove_imp(E,A,_,Vs):-memberchk(E:A,Vs).
\end{code}

\section{Content and order in propositional intuitionistic implicational logic}\label{intu}

Next we will express the ``next token prediction'' as logical inference in implicational intuitionistic logic.
Thus, we would like, from an implicational representation of ``the cat'' and a representation of a sequence of tokens like ``the cat sits'' , to infer the next token ``sits'', by just using {\em modus ponens}. We would also like
the inference steps to be directly verifiable by our theorem provers.

\subsection{Order Preservation by Left-Nested Implication}

Each implication introduces a higher-order dependency on the entire previous structure. Any permutation alters the functional type of the antecedent, changing the set of inhabitants.
Let $w_1,\dots,w_n$ be atomic propositions and define
\[
L_n = ((((w_1 \to w_2) \to w_3) \to \dots) \to w_n).
\]
Then $L_n$ is not invariant under permutation of the $w_i$ in intuitionistic logic.

\BX Counterexample for order invariance in left-nested implication chain.
\begin{codex}
?- iprove(((p->q)->r) -> ((q->p)->r)).
false.
\end{codex}
\EX

By contrast, right-nested implication $R_n$ collapses to a conjunction-to-result form $C_n$
and is $w_1 \ldots w_{n-1}$ permutation-invariant given that conjunction is commutative.

\[
R_n = w_1 \to w_2 \to w_3 \to \dots \to w_{n-1} \to w_n.
\]

\[
C_n = w_1 \land w_2 \land w_3 \land \dots \land w_{n-1} \to w_n.
\]

\BX Order invariance for right-nested implication:
\begin{codex}
  ?- iprove((p->q->r) -> (q->p->r)).
  true.
\end{codex}
\EX

Note that expressing {\em modus ponens}
in terms of implications
 makes it testable with a purely
implicational prover, given the equivalence between
$(p \land (p \to q)) \to q$ and $p \to (p \to q) \to r$.

\subsection{From sequences to implication chains}

Given that our tokenized input is expressed as a Prolog list of words,
we will need to convert this list to left-nested implications.

The predicate {\tt list2impl/2} converts a list to an implication chain
while also handling special cases of empty and single-element lists.
\begin{code}
list2impl([],[]).
list2impl([X],X).
list2impl([X,Y|Xs],R):-seq2left_nested([Y|Xs],X,R).

seq2left_nested([],End,End).
seq2left_nested([X|Xs],Chain,End):-seq2left_nested(Xs,(Chain->X),End).
\end{code}

\BX  Implication chain from list:
\begin{codex}
?- list2impl([the,cat,sits,on,the,map],R).
R = (((((the->cat)->sits)->on)->the)->map).
\end{codex}

\noindent The inverse operation is implemented by {\tt impl2list/2} with
help from DCG-based {\tt impl2list/3 that unfolds the implication chain to a list}.
\begin{code}
impl2list(E,Xs):-impl2list(E,Xs,[]).

impl2list(E->X)-->!,impl2list(E),[X].
impl2list(X)-->[X].
\end{code}
\EX

\subsection{Next-Token Prediction as Modus Ponens}

Given a prefix implication $I_p$ and full implication $I_f$,
\[
I_f = (I_p \to w)
\]
producing $w$ is an instance of modus ponens:
\[
I_f \to (I_p \to I_f) \to w.
\]
Thus, next-token prediction  corresponds to constructive proof completion in intuitionistic implicational logic.
Note that implication depth corresponds to the Modus Ponens inference steps, more precisely
if $L_n = ((((w_1 \to w_2) \to \dots) \to w_n)$ is a left-nested implication chain,
Then $L_n$ requires exactly $n-1$ successive implication introductions and eliminations, as each nesting level introduces one implication whose antecedent is the entire previous chain.

\BX
Given the left-nested implicational representation of the tokens in a sentence
and its infixes, we infer the next word by applying
{\em modus ponens} (reflected to its implicational equivalent):

\begin{codex}
?- iprove(((the->cat)->sits) -> (the->cat) -> sits).
true.
\end{codex}
\EX
\subsection{Stepping deeper into the Curry--Howard interpretation}

Under the Curry--Howard isomorphism, propositions correspond to types,
and proofs correspond to $\lambda$-terms inhabiting those types.
From this perspective, next-token prediction can be viewed as the
construction of a term inhabiting an implicational type determined by
the prefix.

\begin{center}
    \begin{tabular}{lll}
        \toprule
        Intuitionistic Implicational Logic
        & $\lambda$-Calculus (Curry--Howard)
        & Language Modeling \\
        \midrule
        Proposition
        & Type of a $\lambda$-term
        & Token as operator \\

        Implication $A \to B$
        & Function type $A \to B$
        & State transition operator \\

        Proof of $A \to B$
        & $\lambda$-abstraction
        & Learned token operator \\

        Proof term
        & $\lambda$-term
        & Prefix representation \\ 

        Modus ponens
        & Function application
        & Next-token inference \\

        Normalization of proofs
        & $\beta$-reduction
        & Recurrent state update \\
        \bottomrule
    \end{tabular}
\end{center}

In this interpretation, a sentence prefix corresponds to a partially
applied $\lambda$-term whose type is a left-nested implication.
Predicting the next token amounts to supplying an argument that enables
a further function application, i.e., an instance of modus ponens.
The corresponding neural architecture can thus be seen as learning a numeric
realization of proof construction, where recurrent state updates
correspond to successive $\beta$-reductions in the associated
$\lambda$-term.


\section{Theorems (valid formulas) about single next word and multi-word predictions}\label{theo}
A formula is called {\tt valid} if if is an axiom or a theorem in implicational logic,
which, in practice means that it is provable by our theorem provers.

We will illustrate our axioms and theorems as implicational formulas
followed by proof terms computed by {\tt lprove/2}.

\paragraph{The combinator axioms and modus ponens as valid implicational formulas}:

K combinator:
 \verb|p->q->p ==> |
$ \lambda X. \lambda Y.X $

S combinator:
\verb|(p->q->r) -> (p->q)->p->r ==> |
$ \lambda X. \lambda Y. \lambda Z.((X ~Z )~(Y ~Z ))$

Modus Ponens:
\noindent  \verb|p -> (p->q) -> q ==> |
$ \lambda X. \lambda Y.(Y ~X )$

\paragraph{Theorems relevant for single- and multi-word prediction}
We present a few valid implicational formulas (verified with \texttt{lprove/2})
that clarify how a prefix can support both single next-word inference and multi-word
    completion. The unifying idea is that left-nested implication chains naturally induce
continuations: formulas of the form $A \to B$ that, under Curry--Howard, correspond to
functions mapping a proof of $A$ into a proof of $B$. In the language-modeling reading, this
suggests how one may construct a mechanism that extends a prefix one token at a time.

\BX Left nested implies right nested implication chain:

\noindent \verb|((p->q)->r) -> (p->q->r) ==> |
$ \lambda X.\lambda Y.\lambda Z.\big(X~(\lambda U.\lambda V.Z)\big)$
\EX
This theorem expresses \emph{currying}: a context that consumes a composite implication
$(p\!\to\!q)$ to produce $r$ can be transformed into a predictor that consumes $p$ and then $q$ to produce $r$. Note that as \verb|(p->q->r)| and  \verb|(q->p->r)| are equivalent, it also
says that an  order-sensitive left-nested chain is stronger (and more informative) than its permutation-invariant right-nested variant.

\BX Multiple next words :

\noindent  \verb|E -> (((E->a)->b)->c) -> ((a->b)->c) ==> |
$ \lambda X.\lambda Y.\lambda Z.\big(Y~(\lambda U.\lambda V.(Z~(V~X)))\big)$
\EX
This formula provides a schematic link between single-step and multi-step completion.
Given a fixed context $E$ and a ``two-step'' implicational chain from $(E\!\to\!a)$ to $b$ and then to
$c$, we can derive a continuation $(a\!\to\!b)\!\to\!c$ that no longer mentions $E$ explicitly.

For multi-word prediction, this suggests an important pattern:
the prefix context can be used to construct a continuation that generates several future tokens by
iterated application.
Such a continuation is precisely the kind of object that, under our extended Curry--Howard correspondence maps to
composable functions,
indicating how multi-token generation can be framed as repeated implication
elimination.

\section{Information retrieval with logic formulas}\label{info}

When our provers work on the sequent calculus representation of the assumptions
needed to prove a theorem, these assumptions are represented as a list.
This is needed as the proof procedure progressively reduces
this list of assumptions to a simpler form.
As we can reason exclusively with modus ponens on our
left-nested implication chains, we can replace with a fact database
representation this list representation of our assumptions
(each derived from a sentence of a document).
We will add each of the assumptions represented
as a left-nested implication into the dynamic predicate \verb~isent/1~.

To implement an information retrieval mechanism we would use a query matching
a subformula as a retriever of the sentence(s) it originates from.

\subsection{Assuming the subformulas that imply the full sequence representation}

We start by extracting the set of prefixes of the suffixes
of a formula, each such fragment corresponding to a sequence of tokens,
that, when used as a query, should retrieve the formula.

The predicate {\tt ipref/2} extracts the prefix of an implication chain.
\begin{code}
ipref(X,X).
ipref(Xs->_,Ys):-ipref(Xs,Ys).
\end{code}
The predicate {\tt isuff/2} extracts the suffix of and implication chain:
\begin{code}
isuff(X,Y):-isuff_(X,Y).
isuff(X,X).

isuff_(_->X,X).
isuff_(Xs->X,(Ys->X)):-isuff_(Xs,Ys).
\end{code}

Via a DCG notation that composes two relations, we can
generate a prefix of a suffix of implication chain, or equivalently
a suffix of a prefix, giving the same set of solutions,
except for the order in which they are generated.
\begin{code}
isufpref-->isuff,ipref.
iprefsuf-->ipref,isuff.
\end{code}

\BX Results of {\tt isufpref} on a small left-nested formula:
\begin{codex}
?- isufpref((((the->little)->cat)->sits),R).
    R = sits ; R = (cat->sits) ; R = cat ; R = ((little->cat)->sits) ;
    R = (little->cat) ; R = little ; R = (((the->little)->cat)->sits) ;
    R = ((the->little)->cat) ; R = (the->little) ; R = the.
\end{codex}
\EX

To implement a basic information retrieval mechanism we would
like each of these fragments to fetch the corresponding sentence
from our database of implicational formulas representing them.
In Prolog, we could implement these assumptions
as clauses of the dynamic predicate {\tt isent/1}. Note that
if the depth of the left-nested formula is $n$, {\tt isufpref} will generate $n * (n+1) / 2 $ terms.
Thus, to avoid the quadratic increase in size, we could just
compute the \verb~Fragment->Formula~ assumptions on the fly,
done efficiently by the predicate {\tt isufpref/2}, guided
by unification with the query formula.

\subsection{Information retrieval with subformulas as queries}

\subsubsection{Creating the database of implicational formulas}

We first convert each sentence (seen as a Prolog atom) to a left-nested implication chain:
\begin{code}
sent2impl(Sent,R):-
  atomic_list_concat(Words,' ',Sent),
  list2impl(Words,R).
\end{code}
and then we store implication chains of a set of given sentences in the database.
\begin{codeh}
:-dynamic isent/1.
store_impls(Sents):-
  retractall(isent(_)),
  member(Sent,Sents),
    distinct(Impl,sent2impl(Sent,Impl)),
    assertz(isent(Impl)),
  fail;true.
\end{codeh}

\subsubsection{Querying the database of implicational formulas}
The following query-answering predicates implement our
logic-based information retrieval mechanism.

Querying with text string against chains database:
\begin{code}
qa(TextQuery,ISent):-
    string_lower(TextQuery,Query1),
    atom_string(QueryAtom,Query1),
    atomic_list_concat(Words,' ',QueryAtom), 
    lqa(Words,ISent).
\end{code}

We can also query with a list of words some of which could be logic variables
standing for unknown intermediate next words.
\begin{code}
lqa(Words,ISent):-
    list2impl(Words,LeftImplQuery), 
    iqa(LeftImplQuery,ISent).
\end{code}
To find in the database a sentence whose implication chain
has as prefix of a suffix inside a stored implication chain:
\begin{code}
iqa(LeftImplQuery,ISent):-
    isent(ISent), 
    distinct(ISent,isufpref(ISent,LeftImplQuery)).
\end{code}

\begin{codeh}
/*

?- iprove(((p->q)->r) -> (p->q->r)).
true.

?- lprove(((p->q->r)->(p->q)->p->r),SCombinator).
SCombinator = l(_A, l(_B, l(_C, a(a(_A, _C), a(_B, _C)))))

?- lprove(((p->q)->r) -> (p->q->r), LambdaExpr).
LambdaExpr = l(_A, l(_, l(_B, a(_A, l(_, l(_, _B)))))).

?- sent2impl('the cat sits on the mat',Impl).
Impl = (((((the->cat)->sits)->on)->the)->mat).

?- store_impls.
:- dynamic isent/1.

isent((((((the->cat)->sits)->on)->the)->mat)).
isent((((((the->dog)->sits)->on)->the)->log)).
isent(((((the->cat)->chases)->the)->mouse)).
isent(((((the->dog)->chases)->the)->cat)).

?- qa('the cat').
'the cat sits on the mat'

'the cat chases the mouse'

'the dog chases the cat'

?- lqa([the,X,chases],R).
X = cat,
R = ((((the->cat)->chases)->the)->mouse) ;
X = dog,
R = ((((the->dog)->chases)->the)->cat) .

*/
\end{codeh}


\section{The Arrow Model: Neural Realization of Implicational Composition}\label{neural}
\label{sec:neural-realisation}

This section describes a concrete neural realization of the left-nested
implicational semantics introduced earlier.
The goal is not to propose a novel neural architecture per se, but to
demonstrate that a well-defined proof-theoretic interpretation leads
naturally to a recurrent, operator-based model closely related to known
multiplicative RNN architectures.

\subsection{Hidden State as Proof Context}

Our {\em Arrow} model maintains a hidden state
\[
h_t \in \mathbb{R}^d
\]
which represents the accumulated proof context after consuming a prefix
of tokens $x_1,\dots,x_t$.
From a Curry--Howard perspective, $h_t$ is a compact numeric encoding of
the assumptions and justifications constructed so far.

The initial state $h_0$ is a learned vector that seeds the computation at
the beginning of each sequence.

\subsection{Tokens as Implication Operators}

Each token $t$ is treated not as a feature vector to be added, but as an
\emph{operator} that transforms the current proof context.
Formally, each token induces a linear operator
\[
M_t : \mathbb{R}^d \rightarrow \mathbb{R}^d
\]
which is applied to the current state.
To make this feasible for large vocabularies, the operator is
parameterized in low-rank form:
\[
M_t = I + U \, \mathrm{diag}(s_t) \, V^\top,
\]
where:
\begin{itemize}
    \item $U, V \in \mathbb{R}^{d \times r}$ are shared trainable matrices,
    \item $s_t \in \mathbb{R}^r$ is a token-specific gate vector,
    \item $r \ll d$ is a small rank hyperparameter.
\end{itemize}

The gate vector $s_t$ is obtained by embedding lookup and passed through
a $\tanh$ nonlinearity to keep operator perturbations bounded.

This construction yields an operator family that is expressive,
parameter-efficient, and explicitly non-commutative.

\subsection{State Update as Modus Ponens}

The recurrent update implements implication elimination (modus ponens)
as operator application:
\[
h_{t+1}
=
\mathrm{LayerNorm}\!\left(
h_t
+
U\big((V^\top h_t) \odot s_t\big)
\right).
\]
\noindent This update can be read directly as:
    {\em apply the implication introduced by token $t$ to the current proof
    context}.

Because matrix composition is non-commutative in general,
\[
M_a M_b \neq M_b M_a,
\]
the order of tokens is encoded intrinsically by the algebra of operator
composition.
Thus, no explicit positional encoding is required in the baseline model.

\subsection{Relation to Left-Nested Implication}

After consuming tokens $x_1,\dots,x_t$, the hidden state corresponds to
the left-nested implication
\[
((((x_1 \rightarrow x_2) \rightarrow x_3) \rightarrow \dots) \rightarrow x_t)
\]
applied to the initial seed $h_0$.

Note that each recurrence step directly corresponds to one implication elimination.

\subsection{Output Projection and Training Objective}

After each update, the model predicts the next token using a standard
linear readout:
\[
\mathrm{logits}_t = W_{\text{out}} h_t,
\]
where $W_{\text{out}} \in \mathbb{R}^{|\mathcal{V}| \times d}$ projects the
current proof context into vocabulary logits.

Training uses standard next-token cross-entropy with teacher forcing.
Importantly, while the \emph{internal representation of composition}
differs from Transformer models, the \emph{external training objective}
remains unchanged.
This allows direct comparison on the same datasets and metrics.

\subsection{Streaming Loss Computation}

Because the vocabulary may be large (e.g.\ BPE vocabularies with $\sim
10^5$ tokens), computing logits for all time steps simultaneously can be
prohibitively expensive.

Instead, the loss is computed in a streaming fashion:
\begin{itemize}
    \item at each time step, logits of shape $(\text{batch} \times
    |\mathcal{V}|)$ are materialized,
    \item cross-entropy is computed immediately,
    \item logits are discarded before the next step.
\end{itemize}

This reduces peak memory usage and enables training on commodity GPUs
(e.g., 8GB, 16GB or 24GB VRAM) or even m4 Mac Minis with 32GB shared RAM, without specialized kernels.

\subsection{Interpretation and Extensions}

From a logic programming perspective, the Arrow model can be viewed as a
differentiable abstract machine in which each token is an instruction
that mutates a hidden store.
From our extended Curry--Howard perspective, it is a numeric realization of proof
construction by successive implication elimination.

While the baseline model is strictly recurrent and sequential, it can be
extended with lightweight causal mixers or attention layers without
abandoning the core implicational semantics.
Such hybrids are left for future work.

We refer to our online
code\footnote{\url{https://github.com/ptarau/nextword/blob/main/arrow.py}}
for the details of the implementation of training and inference.

\section{Experiments}\label{exper}

Our experiments are organized around a guiding hypothesis:
\emph{left-nested implicational structure provides an order-sensitive representation
    that supports both symbolic retrieval (Section~\ref{info}) and neural next-token
    prediction via operator composition}.
To make the link between the logic-level retrieval mechanism and the learned
neural mechanism maximally transparent, we use a deliberately overfitted training
regime that turns the training corpus into a ``memorized'' store of sentences and
sentence fragments. This mirrors the Prolog dynamic database of assumptions used
by the logic retrieval procedure.
Attempting to overfit a small document or batch—often called a "Sanity Check"—is a common and effective indicator for evaluating a transformer (or similar) neural architecture choice.
While standard overfitting is typically viewed negatively, in this specific context, it serves as a diagnostic tool for things like verification of learning capacity, identifying overall architectural efficiency and
normalization check to identify if specific layers are correctly handling the scale of the data.


\paragraph{Gutenberg document acquisition.}
We obtain public-domain  texts from Project Gutenberg\footnote{
\url{https://gutenberg.org}} and treat each document as a single large corpus.
For each document, we automatically download the plain-text UTF-8 variant,
remove the standard Gutenberg header/footer boilerplate, and retain
only the main body. This yields large, clean corpora without licensing restrictions.

\paragraph{Sentence sets and normalization.}
Each document is converted into a set of sentences, with one sentence per line,
to support two complementary pipelines:
(i) conversion into implicational formulas for Prolog-based retrieval, and
(ii) tokenization for neural training and evaluation.
We apply lightweight normalization (lowercasing and punctuation cleanup) to
reduce accidental duplication caused by formatting artifacts.
The resulting sentences serve as the \emph{ground truth store} for both Prolog
and Python retrieval experiments.


\paragraph{Fragments as implicational subformulas.}
Section~\ref{info} shows that retrieval in the logic setting can be driven by a
query that is a \emph{suffix subformula} (or prefix of such a suffix) of a stored
sentence formula. In the left-nested encoding, a contiguous token span
corresponds to a syntactically identifiable substructure. To mimic this in the
neural setting, we augment the training set with \emph{all contiguous fragments}
of each sentence, i.e., all subsequences $x_i,\dots,x_j$ that appear contiguously
in a sentence $x_1,\dots,x_n$.

Concretely, for each sentence token sequence $(x_1,\dots,x_n)$ we include all
fragments $(x_i,\dots,x_j)$ for $1 \le i \le j \le n$.
We deduplicate fragments globally across the document to avoid bias from repeated
lines (common in large novels and dialogue-heavy texts) and set a  practical limit
on the subformula size used in the training loop.

\paragraph{Our parallel symbolic and neural retrieval}

In the Prolog pipeline, each sentence is stored as an implicational formula in a
dynamic database of assumptions, and queries are posed as subformulas (or prefixes
of suffix subformulas) of these stored formulas.
Retrieval succeeds when the query matches a stored sentence structure under the
intended implication encoding; this corresponds to proving or reconstructing the
completion using intuitionistic principles described earlier.
This provides a crisp, interpretable baseline: \emph{retrieval is exact and
    structure-driven}.

The Python implementation mimics the logic retrieval process in
two stages:

\emph{(1) Candidate enumeration by subsequence match.}
Given a user query consisting of a short word sequence $q=(q_1,\dots,q_m)$, we
scan the stored sentences (the same sentence set used for training) and collect
all occurrences where $q$ appears as a contiguous subsequence. Each occurrence
induces a candidate \emph{suffix completion} (from the match position to the end
of the sentence). This mirrors the way a suffix subformula of a stored
implicational chain determines a unique completion in the symbolic setting.

\emph{(2) Ranking by learned continuation score.}
For each candidate suffix, we compute a log-likelihood score under the trained
Arrow model for the continuation \emph{after} the matched query. Intuitively,
this score measures how compatible the candidate completion is with the learned
operator dynamics induced by the query prefix.
This neural retrieval mode can be seen as a ``soft'' analogue of the Prolog
retrieval: instead of logical entailment/unification, we use learned conditional
probability to select the most plausible completion among exact substring matches.

\paragraph{Performance considerations}
Both approaches scale in a predictable way.

     \emph{Prolog retrieval} scales with the size of the dynamic database
    and the indexing strategy used for matching subformulas. Because queries are
    structural, retrieval is typically fast once the database is built. Performance
    can be fine-tuned with indexing on leading functors/atoms or via {\tt term\_hash/2},
    given that the terms involved are ground.

     \emph{Neural scoring} scales linearly in sequence length per candidate
    and it benefits substantially from GPU acceleration.
    Our use of mixed precision and streaming loss also reduce memory pressure,
    allowing training and inference on a single commodity GPU.
    In the Arrow model, the
    per-step recurrence is inexpensive and scales linearly with the
    maximum length of the sentences.

    On the other hand, the number of sentences
    and the number of words have a limited impact on the training time,
    given that the batched training operations are easily parallelized on the GPU.

\subsection{Scaling and Runtime Measurements}
\label{sec:runtime}

\paragraph{Experimental plan for large-scale runs.}
For large documents we report corpus statistics
(number of sentences, vocabulary size, number of unique
fragments under deduplication),

All experiments in this section use a maximum sentence length of $256$ words and an
augmentation regime that includes all contiguous fragments up to length $5$ (globally deduplicated).
This mirrors the logic-level setting of Section~\ref{info}, where queries correspond to subformulae
of stored implicational chains, and it matches the Python retrieval pipeline, where a query is a
contiguous subsequence whose best suffix completions are ranked by a learned continuation score.

\paragraph{Training throughput and scaling.}
Under the above fixed constraints, training completes in a few minutes per document
(Table~\ref{tab:train-times}).

\begin{table}[t]
    \centering
    \small
    \begin{tabular}{lrrrll}
        \toprule
        \textbf{Document} & \textbf{Sentences} & \textbf{Words} & \textbf{Training time} & \textbf{Notes} \\
        \midrule
        \texttt{the\_eyes}        &    94 &   1121  & 6m 47.355s & 256-word sentences, fragments $\le 5$ \\
        \texttt{war\_and\_peace}  & 29318 & 560285  & 7m 52.297s & same settings \\
        \texttt{ulysses}         & 17564 & 254917  & 9m 12.960s & same settings \\
        \texttt{guermantes}      &  7848 & 260740  & 6m 43.017s & same settings \\
        \texttt{wizard\_of\_oz}   &  1158 &  43092  & 4m 20.580s & same settings \\
        \texttt{moby\_dick}       &  8728 & 213170  & 6m 47.355s & same settings \\
        \texttt{dracula}         &  6150 & 161410  & 5m 20.936s & same settings \\
        \texttt{cthulhu}         &   364 &  12074  & 4m 43.505s & same settings \\
        \texttt{crystal}         &   369 &   7472  & 5m 54.452s & same settings \\
        \bottomrule
    \end{tabular}
    \caption{End-to-end training runtimes on several Gutenberg documents under a fixed preprocessing
        and augmentation protocol: maximum sentence length $256$ words and maximum fragment length $5$
        on a Linux machine with 32GB RAM and an RTX 3090 GPU with 24GB VRAM.}
    \label{tab:train-times}
\end{table}

\paragraph{Inference-time behavior.}
Inference in the retrieval-first Python pipeline (candidate enumeration by subsequence match
followed by continuation scoring) takes approximately $0.1$--$0.3$ seconds per query, and was observed
to be largely independent of the overall file size in these experiments.
This is consistent with the fact that, once candidate suffixes are identified, scoring requires only
short sequential evaluation over the candidate continuation (bounded by the maximum sentence length),
rather than processing the entire document.

\paragraph{Prolog database construction and query time.}
In the Prolog baseline, loading the dynamic database of sentence formulas from a preprocessed
sentence file takes approximately $1$--$2$ seconds, while typical queries complete in roughly
$0.5$ seconds.
This supports the feasibility of using logic-level retrieval as an interpretable baseline even for
large documents: database initialization is fast, and query evaluation remains interactive.

\section{Discussion}\label{disc}
\paragraph{Limitations.}
The present work deliberately focuses on settings where the query/prefix is an \emph{exact} contiguous
substructure of a stored sentence (or fragment), so that both the Prolog retrieval procedure and the
Python mimic operate under exact matching assumptions. While this controlled regime is useful for
making the implicational semantics explicit, it does not address the robustness requirements of
modern language modeling, such as prompts containing misspellings, omissions, paraphrases, or
non-contiguous evidence.

\paragraph{Future work directions: towards a proof-theoretic account of transformers}
A central open challenge is to emulate and explain transformer-style inference using similar
logic-based techniques. In particular, transformers can perform next-token prediction even when the
prompt is noisy or partially specified; this suggests an inference mechanism that behaves like
\emph{approximate} matching and \emph{soft} hypothesis selection rather than exact unification.

\section{Related Work}\label{rel}

\paragraph{Logic Programming and Neuro-Symbolic works related to LLM Reasoning}
A growing line of neuro-symbolic work treats an LLM primarily as a \emph{semantic parser} that maps natural language into a formal language (FOL, ASP, Prolog), and then delegates multi-step reasoning to a symbolic engine with explicit proof traces,
\cite{olausson2023linc,yang2023coupling}
In \cite{flops24} deep step-by step reasoning in an LLM dialog thread is automated by recursively exploring alternatives (OR-nodes) and expanding details (AND-nodes) up to a given depth.
The algorithm is derived from a simple recursive descent implementation of a Horn Clause interpreter. Semantic similarity to ground-truth facts or oracle advice from another LLM instance is used to restrict the search space and validate the traces of justification steps returned as focused answers.
In the ASP direction, Ishay et al.\ \cite{ishay2023asp} show that LLMs can generate nontrivial ASP encodings for logic puzzles (often with simple, human-correctable errors), enabling a workflow where correctness is enforced by a solver rather than implicit pattern completion.

These approaches are complementary to ours: rather than outsourcing deduction to an external prover/solver, we aim to \emph{internalize} an implication-shaped constructive update in the neural dynamics, while retaining a semantic link to proof-theoretic structure.


\paragraph{Comparison with related neural models}


A standard Transformer \cite{transfo} represents tokens as embeddings added into a
residual stream and relies on self-attention to retrieve and mix past
information.
Order is injected primarily through positional encodings or rotary
embeddings.

In contrast, our Arrow model:
\begin{itemize}
    \item represents tokens as state-transforming operators,
    \item encodes order through non-commutative composition,
    \item stores history implicitly in the evolving state rather than
    via content-addressable retrieval.
\end{itemize}

This shifts the inductive bias from retrieval to transformation.
A token influences future predictions not by being attended to, but by
having permanently modified the proof context.

A large body of work refines positional structure while keeping attention as the core interaction \cite{DBLP:journals/ijon/SuALPBL24,DBLP:conf/iclr/PressSL22}.
These methods emphasize that, in the transformer family, order is primarily enforced by
positional augmentation of an otherwise similarity-driven (dot-product) interaction.

Our model depart from this design choice: rather than adding positional information to a
similarity operator, we encode order directly via non-commutative composition.
A prefix is interpreted as a left-nested implicational chain acting on a state,
so the model's sequential backbone is intrinsically order-sensitive.
In this sense, the Arrow model shifts the inductive bias from \emph{positionalized similarity}
to \emph{operator composition}.


State-space sequence models (SSMs) provide an alternative to attention in which long-range
dependencies arise from structured linear dynamics \cite{gu2024mambalineartimesequencemodeling} .
Our Arrow model differs from classical SSMs in that its recurrence is explicitly \emph{token-dependent} and
operator-valued, rather than governed by a fixed global transition/kernel.
This places Arrow closer to the family of input-conditioned recurrent models, while sharing the
SSM motivation of making long-context computation efficient and structurally grounded.


The Arrow model update
is mathematically closest to {\em recurrent models in which the input chooses a
state transition operator} like
\cite{suts11} using a multiplicative RNN for language modeling in which
the current input modulates the hidden-to-hidden transformation via a low-rank factorization
to show strong results when combined with Hessian-free optimization.

In  \cite{zhang2025rlm}  \emph{Recursive Language Models} (RLMs) are introduced as an inference-time strategy
for handling prompts far longer than the model’s context window, by treating the prompt as an external
environment that the model can examine, decompose, and recursively re-query.
In \cite{tarau2023automation}  a logic-programming inspired method for \emph{automating goal-driven LLM dialog
    threads} by organizing reasoning as a bounded-depth recursive exploration of OR-alternatives and AND-expansions,
explicitly derived from Horn-clause (SLD-resolution-like) control.
 This work complements our approach by showing how Logic Programming control principles can structure and
validate LLM inference externally, whereas our present paper focuses on how implicational structure
itself motivates an internal, compositional view of next-token prediction and multi-token continuation.

\section{Conclusion}\label{conc}

We presented the Arrow Language Model, grounded in \emph{left-nested} intuitionistic implication.
A formal correspondence links valid  implicational formulas
to neural recurrent computation, explaining why order
is intrinsic to the architecture. This yields a logic-first foundation for next-token prediction
distinct from similarity-based attention and from convolution/state-space alternatives.
For the reader curious to explore independently our open source code, a quick way to
start are the simple commands shown at \url{https://github.com/ptarau/nextword/blob/main/README.md} .

\bibliographystyle{eptcs}

\bibliography{tarau,llm,ml,proglang, theory}

\end{document}